\documentclass[aps,prl,showpacs,twocolumn,superscriptaddress,nofootinbib,groupedaddress]{revtex4} 
\usepackage{color}
\usepackage{graphicx}
\usepackage{amsmath}
\usepackage{braket}
\usepackage{amssymb}
\usepackage[colorlinks=true,citecolor=darkred,urlcolor=darkred, pdfborder={0 0 0}]{hyperref}
\definecolor{darkred}{rgb}{0.6,0,0}

\definecolor{linkcolor}{rgb}{0,0,0.5}

\def\gsim{\raise0.3ex\hbox{$\;>$\kern-0.75em\raise-1.1ex\hbox{$\sim\;$}}}
\def\lsim{\raise0.3ex\hbox{$\;<$\kern-0.75em\raise-1.1ex\hbox{$\sim\;$}}}

\def\beqn#1{\begin{equation}\label{#1}}
\def\eeqn{\end{equation}}

\def\beqa#1{\begin{eqnarray}\label{#1}}
\def\eeqa{\end{eqnarray}}

\def\Z2{$\mathcal{Z_2}$}

\def\vev#1{\left\langle #1\right\rangle}

\newcommand{\sm}{{Standard Model }}
\newcommand{\AddrAHEP}{%
  AHEP Group, Institut de F\'{i}sica Corpuscular --
  C.S.I.C./Universitat de Val\`{e}ncia, Parc Cient\'ific de Paterna.\\
 C/ Catedr\'atico Jos\'e Beltr\'an, 2 E-46980 Paterna (Valencia) - SPAIN}

\def\SMB-L{$\mathrm{SU(3)_c \otimes SU(2)_L \otimes U(1)_Y\otimes U(1)_{B-L}}$ }

\begin{document}

\title{A Model of Comprehensive Unification}
 
\author{Mario Reig} \email{mario.reig@ific.uv.es}
\affiliation{\AddrAHEP}

\author{Jos\'e W.F. Valle} \email{valle@ific.uv.es}
\affiliation{\AddrAHEP}

\author{C.A. Vaquera-Araujo} \email{carolusvaquera@gmail.com}
\affiliation{\AddrAHEP}

\author{Frank Wilczek} \email{wilczek@mit.edu} \affiliation{Center for
  Theoretical Physics, MIT, Cambridge MA 02139 USA}
\affiliation{Wilczek Quantum Center, Department of Physics and Astronomy, Shanghai Jiao Tong University, Shanghai 200240, China}
\affiliation{Department of Physics, Stockholm University,  Stockholm SE-106 91 Sweden}
\affiliation{Department of Physics and Origins Project, Arizona State University, Tempe AZ 25287 USA}

\date{\today}



\begin{abstract}
 
  Comprehensive - that is, gauge {\it and\/} family - unification
  using spinors has many attractive features, but it has been
  challenged to explain chirality.  Here, by combining an orbifold
  construction with more traditional ideas, we address that
  difficulty.  Our candidate model features three chiral families and
  leads to an acceptable result for quantitative unification of
  couplings.  A potential target for accelerator and astronomical
  searches emerges.

\end{abstract}

\maketitle

 \section{Introduction}
 
 Our core theory of fundamental physics, based on promoting
 $SU(3)\times SU(2) \times U(1)$ to a local symmetry, describes a vast
 range of phenomena precisely and very accurately.  In that
 sense, it is close to Nature's last word.  On the other hand, it
 contains a diversity of interactions and, when we come to the
 fermions, a plethora of independent elements.  It is attractive to
 imagine that a deeper unity underlies this observed multiplicity.
 Gauge unification, perhaps most elegantly realized using the group
 $SO(10)$ and the spinor {\bf 16} representation of
 fermions~\cite{Georgi:1974sy}, goes a long way toward that goal.  It
 leaves us with a single interaction (i.e., a simple gauge group) but
 three fermion families, each embodying a chiral spinor {\bf 16}
 representation.  It is then natural to ask, whether one can take that
 success further, to unite the separate families.

 The mathematical properties of spinor representations are suggestive
 in this regard \cite{Wilczek:1981iz,Bagger:1984rk}.  Specifically,
 for example, the irreducible spinor {\bf 256} representation of
 $SO(18)$ reduces, upon breaking $SO(18) \to SO(10) \times SO(8)$,
 according to
 ${\bf 256} \to ({\bf 16, 8})~+~(\overline{\mathbf{ 16}}, {\mathbf{
     8^\prime}})$, involving spinor representations of the smaller
 groups (including conjugate and alternate spinors).  From the
 standpoint of $SO(10)$, then, we have eight families and eight mirror
 families.  Notably, there are no problematic exotic color or charge
 quantum numbers: we get basically the sorts of representations we
 want, and no others.  Still, there are too many families, and the
 mirror families carry the ``wrong'' chirality for low-energy
 phenomenology \cite{Fonseca:2015aoa}.  Confinement of some $SO(8)$
 quantum numbers, or interaction with condensates, can effectively
 remove an equal number of families and mirror families, but it seems
 difficult to change their net balance by those means.

 The idea of comprehensive unification has continued
   to atract attention over the years, both in context of $SO(18)$ and
   in variant forms \cite{Babu:2002ti,Barr:1987pu}, but the issue of
   chirality has remained salient.

   In this letter we explore a different direction.  We use an
   orbifold construction to break $SO(18) \to SO(10) \times SO(8)$,
   with chiral fermion zero modes in $ ({\bf 16, 8})$.  In addition we
   postulate condensates that break $SO(8) \to SO(5)$ and decompose
   $ {\bf 8} \to 3\times {\bf 1} + {\bf 5}$.  The $SO(5)$ then becomes
   strongly coupled and confining at a scale
   $\mathcal{O}(\text{TeV})$, effectively leaving three chiral spinors
   of $SO(10)$ at low energies.  When one includes contributions from
   the required Higgs fields, an acceptable fit to gauge coupling
   unification emerges (despite the absence of low-energy
   supersymmetry).  An interesting consequence of this scheme is the
   existence of stable $SO(5)$ hyperbaryons, protected by a $Z_2$
   symmetry.  Although they annihilate in pairs, a significant relic
   density emerges from big bang cosmology.

\section{Model construction}

We will exploit the possibility to obtain chiral fields by imposing
appropriate boundary conditions on orbifolds. That mechanism has been
used, for example, in a recent higher-dimensional extension of the
Standard Model \cite{Chen:2015jta}.  

Supersymmetry will play no role
in our discussion. In the context of warped extra dimensions, a major
motivation for supersymmetry is that it avoids Planck scale radiative
corrections, that would re-introduce the hierarchy problem, when
scalar fields are allowed to propagate in the bulk
\cite{Gherghetta:2000qt}.   Our scalars will be localized on the
branes.  As will emerge below, it is not implausible that we can fulfill the main quantitative
motivation for low-energy supersymmetry - the unification of couplings
- in a different way.  One can, of course, assume that supersymmetry
is present in a more basic underlying theory, but broken at the Planck
scale.  Here, however, we will not address issues of ultraviolet
completion.

Our model employs an $S_1/(Z_2\times Z_2^\prime)$ orbifold.  Specifically, we consider a
circular fifth dimension of radius $R=2L/\pi$, with walls at $y=0, L$
and a warped metric \cite{Randall:1999ee}:
\begin{equation}
  ds^2=e^{-2\sigma(y)}\eta_{\mu\nu}dx^\mu dx^\nu+dy^2\,,
\end{equation}
with
\begin{eqnarray}
 \sigma(y)=\sigma(y+2L)=\sigma(-y) \\ \nonumber
  \sigma(y)=ky\,\,\,\text{for}\,\,\, 0\leq y\leq L\,.
\end{eqnarray}
We define the equivalence relations \cite{Hebecker:2001wq}
\begin{equation}\begin{split}\label{equiv}
    &{\bf P}_0:y\sim -y\,,\\& {\bf P_1} : y^\prime\sim -y^\prime\,.
  \end{split}\end{equation}
where $y^\prime \equiv y + L$ Thus the second relation in
Eq.~(\ref{equiv}) is equivalent to $y\sim y+2L$.  In the standard
Randall-Sundrum terminology, we can say that the bulk region, $0<y<L$,
is sandwiched between a Planck brane ($y=0$) and a IR brane ($y=L$).

The action of these equivalences ${\bf P}_0$, ${\bf P}_1$ on matter fields is
\begin{equation}\begin{split}
    &\Phi (x,y)\sim P_0^\Phi \Phi (x,-y)\,,\\& \Phi (x,y^\prime)\sim
    P_1^\Phi \Phi (x,-y^\prime)\,,
  \end{split}\end{equation}
where $P^\Phi_0$ and $P^\Phi_1$ are matrices that represent the
action of the $Z_2$ on the bulk fields. We can classify fields
by their $(P^\Phi_0,P^\Phi_1)$ values.
It will be convenient to write the orbifold conditions for gauge
  fields as:
\begin{equation}\label{gauge_bc}    \left ( \begin{array}{c}
               A_\mu\\
               A_y
             \end{array}\right)
           (x,y_j-y) \sim P^A_j \left ( \begin{array}{c}
                                  A_\mu\\
                                  -A_y
                                \end{array}\right )
                              (x,y_j+y)(P^A_j)^{-1}\,                        
\end{equation}
where $(y_0, y_1) \equiv (0, L)$.  
Thus 
\begin{equation}
A_{M} (x, y+2L)=UA_{M} (x,y)U^{-1}
\end{equation}
with $U=P^A_{1}P^A_{0}$.

We will choose
\begin{equation}\begin{split}
&P^A_0=\text{diag}(\mathbb{I}_{10},-\mathbb{I}_8)\,,\\&
P^A_1=\text{diag}(\mathbb{I}_{18})\,.
\end{split}\end{equation} 
and the corresponding representation matrices for $P^\Phi_j$.
These boundary conditions reduce $SO(18) \to SO(10) \times SO(8)$.
                          
We can decompose a generic five-dimensional field as:
\begin{equation}
  \Phi(x,y)=\frac{1}{\sqrt{L}}\sum^\infty_{n=0}\phi^{(n)}(x)f_n(y)\,,
\end{equation}
where $\phi^{(n)}$ are the Kaluza-Klein (KK) excitations and the
KK eigenmodes, $f_n(y)$, obey:
\begin{equation}
  \frac{1}{L}\int dy\,e^{(2-s)\sigma}f_m (y) f_n  (y)=\delta_{mn}\,,
\end{equation}
where $s=2,4,1$ when the field is a vector field, a scalar or
a fermion, respectively~\cite{Gherghetta:2000qt}.

In more detail, according to Eq.~(\ref{gauge_bc}), the $SO(18)$ gauge
adjoint representation will split as
\begin{equation}
\mathbf{153}=(\mathbf{45},\mathbf{1})^{++}+(\mathbf{1},\mathbf{28})^{++}+(\mathbf{10},\mathbf{8})^{-+}\,,
\end{equation}
so only adjoint fields corresponding to $SO(10)\times SO(8)$ have zero
modes. Because the fifth components, $A_y$, have opposite boundary
condition, they have only Kaluza-Klein modes.

A left-handed fermion field will have a massless zero-mode only when it has
Neumann $(+)$ boundary conditions at both Planck and IR branes
\begin{equation}
  \phi^{(++)}(x,y)=\frac{1}{\sqrt{L}}(\phi^{(0)}_{++}(x)f(y)^{(0)}+\text{higher modes})\,,
\end{equation}
The same occurs with right-handed fields that have Dirichlet $(-)$
boundary conditions at both branes, while fields with $(+,-)$ or
$(-,+)$ do not have zero modes regardless of their chirality. The
$\phi^{(0)}(x)$ zero mode is a massless field in four
dimensions, while the $\phi^{(n)}(x)$ Kaluza-Klein modes have masses
of order $\mathcal{O}(1/L)$, and do not appear in the low-energy
spectrum of the theory.

For the fermion spinor we have~\cite{Slansky:1981yr}:
\begin{equation}\label{BC}\begin{split}
    \mathbf{256}=(\mathbf{16},\mathbf{8})^{++}+(\overline{\mathbf{16}},\mathbf{8}^\prime)^{-+}\,.
\end{split}\end{equation}
Since only the first of these supports zero modes, the mirror families
decouple from low-energy phenomenology.

Together with the bulk spinor and gauge fields, we will incorporate
brane-localized scalars which implement spontaneous symmetry breaking
by condensation (Higgs mechanism).  Further breaking to the \sm might
proceed through intermediate steps associated with either a
Pati-Salam~\cite{pati:1974yy} or left-right
symmetric~\cite{mohapatra:1981yp} stage.  However, here we assume just
the simplest case of direct breaking by Higgs fields in the representations
\begin{equation}
  (\mathbf{210},\mathbf{1})+(\mathbf{126},\mathbf{1})+(\mathbf{10},\mathbf{1})\,.
\end{equation}
While the scalars $(\mathbf{210},\mathbf{1})$ and
  $(\mathbf{126},\mathbf{1})$ are localized at the Planck brane, the
  $(\mathbf{10},\mathbf{1})$ is confined to the IR brane.
Quantitative unification of couplings roughly supports this simplest
choice {\it a posteriori}, as will appear.  The
$(\mathbf{10},\mathbf{1})$ lies at the TeV scale and drives
electroweak breaking.  Planck brane scalars naturally acquire large
masses, thanks to the warp factor.

A  special feature  of $SO(8)$  is  the existence  of three  different
8-dimensional representations:  vector, spinor, and  alternate spinor.
They  are   equivalent  to  one   another  under  a   symmetric  $S_3$
``triality'' group of  outer automorphisms.  For our  purposes, it may
be  simplest to  regard  the spinor  {\bf  8} of  our  fermions as  an
equivalent vector, and break $SO(8) \to SO(5)$ by means of an adjoint,
or three vectors. Alternatively, we might take the spinor as it comes,
and         note         that         it         decomposes         as
${\bf 8}  \to 2 \times  {\bf 1} + {\bf  6}$ under the  natural $SU(4)$
subgroup  of $SO(8)$.   We can  break to  that using  a spinor.   Then
exploiting the isomorphism $SU(4) \to SO(6)$, we break down to $SO(5)$
using   a   vector   of   $SO(6)$.     In   either   case,   we   have
${\bf 8}  \to 3  \times {\bf 1}  + {\bf 5}$  under $SO(8)  \to SO(5)$.
Assuming that  this breaking  occurs through $SO(10)$  singlet scalars
localized on the Planck brane, the details do not influence low energy
phenomenology.

The upshot is that our low-energy fermions transforms as
$3 \times ({\bf 16}, {\bf 1}) + ({\bf 16}, {\bf 5})$ under
$SO(10)\times SO(5)$.  Running of couplings down to low energies
suggests that the $SO(5)$ becomes strongly interacting at
$\mathcal{O}(\text{TeV})$\footnote{Note that if confinement takes
  place above EW scale the only allowed condensate is formed by the SM
  singlet contained in the $(\mathbf{16},\mathbf{5})$.}.  Thus the
${\bf 5}$ will be confined, and at low energies we arrive at just
three chiral spinor families of $SO(10)$, as desired.  (The mechanism
of ``heavy color confinement'' has a long history in this context, see
Refs.~\cite{Wilczek:1981iz,GellMann:1980vs,Giudice:1991sz}).

Proton decay is potentially very rapid, if the scale
  of the IR brane is low.  The simplest solution is to make that scale
  large, e.g. associated with conventional unification or with
  gravitational physics.  In this scenario, we are using the extra
  dimension to address chirality, rather than the hierarchy problem.
  Other solutions may be possible
  \cite{Pomarol:2000hp,Bando:1999di}. 
\section{Gauge coupling evolution }
\begin{figure}[h]
  \centering
  \includegraphics[width=0.45\textwidth]{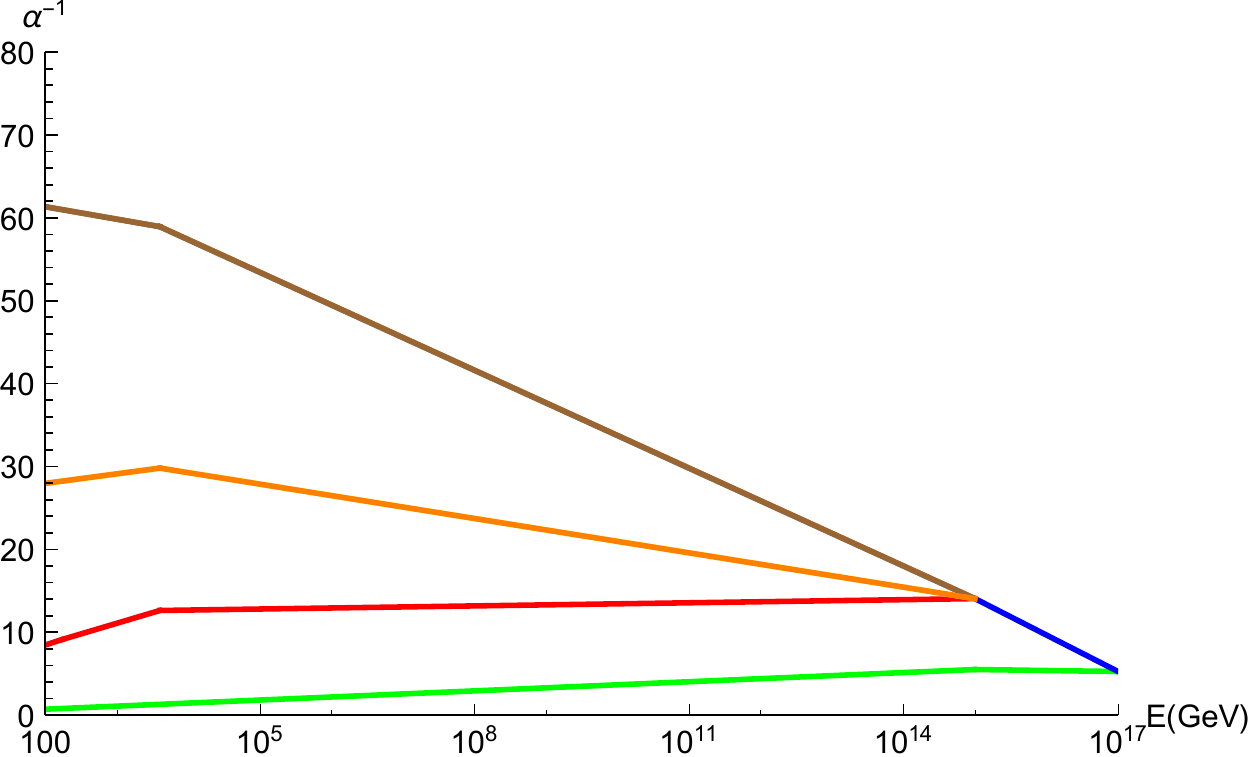}
 \caption{(Color online) Running of gauge couplings (top-down approach): below the SO(10)
    scale we have the SO(5) gauge coupling (green line) in addition to
    the \sm couplings (red, orange and brown lines).  See text.}
  \label{run18}
\end{figure}

One can write the running of the gauge coupling constants in the four
dimensional unified gauge theory
\begin{equation}\label{rge}
\alpha^{-1}_{i}(M_Z)=\alpha^{-1}_{GUT}+\frac{b_i}{2\pi}\log\frac{M_{GUT}}{M_Z}+\Delta_i\,,
\end{equation}
where $\Delta_i$ denote threshold corrections.
Within a five dimensional warped space-time one should take into
account contributions from the Kaluza-Klein modes, as well.  In
\cite{Randall:2001gc} it was argued that warped extra dimensions,
unlike flat extra dimensions, lead to logarithmic running of
couplings. Indeed, an equation similar to Eq.~(\ref{rge}) holds, with
the $b_i$ given as~\cite{Pomarol:2000hp,Randall:2001gc}:
\begin{equation}\begin{split}
b_i^{RS}&=\frac{1}{3}[-C_2(G)(11I^{1,0}(\Lambda)-\frac{1}{2}I^{1,i}(\Lambda))+\\&
+2I^{1/2,0}(\Lambda)T_f(R)+I^{2,0}(\Lambda)T_s(R)]\,.
\end{split}\end{equation}
We take the cut-off scale to be $\Lambda\sim k$, which implies the
numerical values \cite{Randall:2001gc}:
\begin{equation}\begin{split}
&I^{1,0}=1.024 \,,\\&
I^{1,i}= 0.147 \,,\\&
I^{1/2,0}=1.009  \,,\\&
I^{2,0}= 1.005 \,.
\end{split}\end{equation}
For scalars localized on branes, we just change $I^{2,0}(\Lambda) \to 1$.
In Fig.~(\ref{run18}) we fix, for definiteness, the unification
  scale at 10$^{15}$ GeV, and perform a first estimate of the
  electroweak mixing angle within a top-down approach. We find
$\sin ^2\theta_w\approx 0.215$, to be compared with the observed value
$0.22$.  Given our neglect of (inherently uncertain) threshold
corrections and higher order renormalization, this seems an acceptable
result (see below).
\begin{figure}[h]
	\centering
	\includegraphics[width=0.45\textwidth]{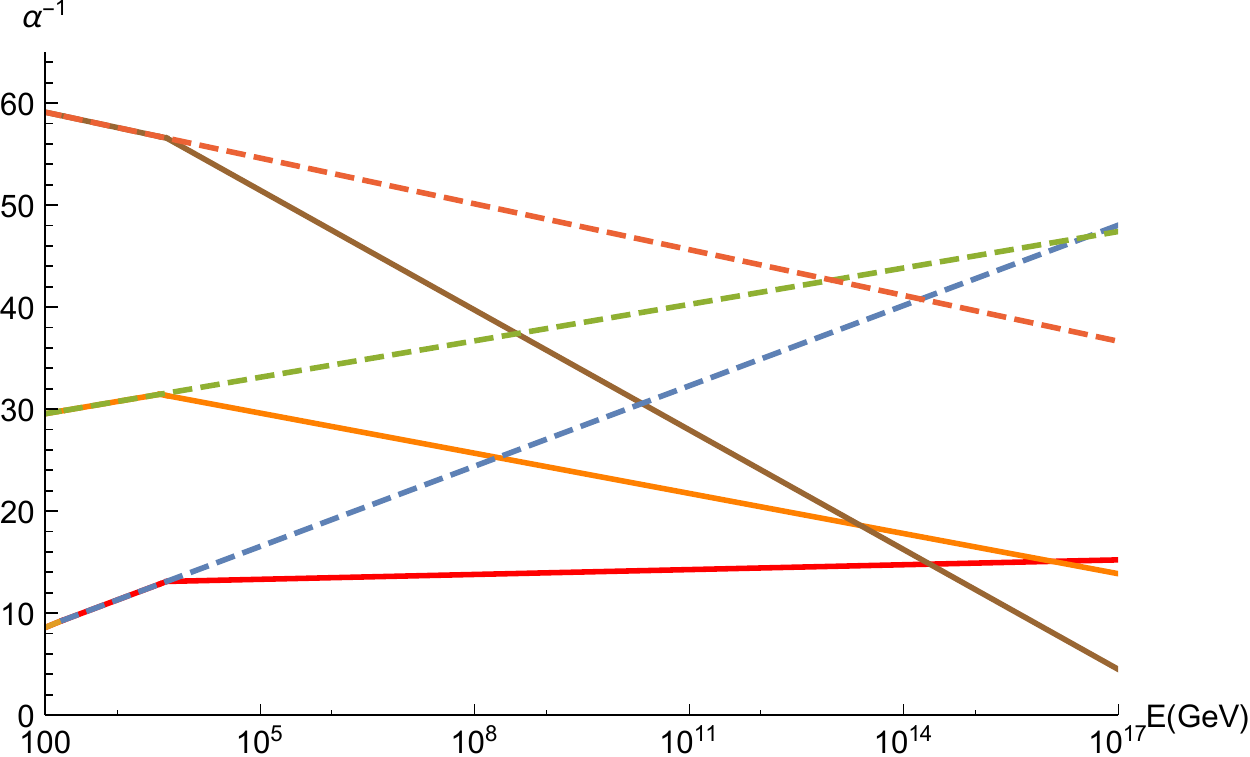}
	\caption{(Color online) Running of gauge couplings below the SO(10)
		scale compared with the SM (dashed lines). Bottom-up approach.}
	\label{sm_like_plot}
\end{figure}

Note that in this simplest case one breaks the $SO(10)$ directly to
$SU(3)\times SU(2) \times U(1)$.  One finds that the SO(5)
coupling reaches non-perturbative values at $\mathcal{O}(\text{TeV})$
(green curve). This fact is reflected into a kink in the
  evolution of the \sm couplings at this value. Thanks to the large
  number of ``active'' flavors, the evolution of $g_3$ is nearly flat
  all the way from few TeV up to the GUT scale (see red curve).
Above the GUT scale $\alpha_{10}$ (blue curve) rises again due to the
large Higgs boson multiplets. 

In Fig.~(\ref{sm_like_plot}) we compare the bottom-up running at one
loop compared with a similar \sm extrapolation.  One sees that our
simple unification scenario gives a marginal improvement with respect
to the minimal \sm case. However, these results come from a rough
estimate, taking renormalization group evolution to first order and
neglecting threshold corrections.

Charged fermion masses arise from the $\vev{(\mathbf{10},\mathbf{1})}$
vacuum expectation value\footnote{The $\mathbf{10}$ scalar belongs to
  a $\mathbf{18}$ localized at the IR brane, where the SO(18) is not
  broken by boundary conditions. When orbifold breaking takes place
  this scalar splits as $\mathbf{18}\to \mathbf{10}+\mathbf{8}$, and
  the $\mathbf{8}$ can be decoupled using a generalized
  Dimopoulos-Wilczek mechanism.}, while neutrino masses can be
induced by the conventional (high scale) seesaw
mechanism~\cite{GellMann:1980vs,glashow1980future,Yanagida:1979as,Schechter:1980gr,Lazarides:1980nt,mohapatra:1981yp}.
Note also that the doublet-triplet splitting problem may be solved
with a generalization of the Dimopoulos-Wilczek mechanism
\cite{Wilczek-Dimopoulos} for $SO(18)$, using a heavy bulk scalar that
leaves the SU(2) doublet massless.

As a final comment we note that the breaking of the
  $SO(3)$ subgroup of $SO(8)$ will be important in connection with the
  flavor puzzle, and could lead to new ways of addressing details of
  the family mass hierarchy and mixing
  pattern. 
  The implementation of specific mechanisms, however, lies beyond
  the scope of our minimal scenario.

\section{Hypercolor and hyperbaryons}

The evolution of each of the SO(10) and SO(8) coupling constants can
be computed imposing the initial unification condition
\begin{equation}
  g_{10}(M_{18})=g_{8}(M_{18})\,,
\end{equation}
at some scale $M_{18}\lsim M_P$ where gauge couplings meet.
(In our concrete estimates we set $M_{18}$, the scale which breaks
$SO(8)$  to $SO(5)$, at $\approx 10^{17}$ GeV.)
The value of $g_{10}(M_{18})$ can be inferred from the observed value
of Standard Model couplings.  The largest Standard Model coupling at
low energies is the $g_3$ of strong $SU(3)$.  Being a larger gauge
symmetry, our $SO(5)$ is ``more asymptotically free'' than $SU(3)$,
and we expect that its coupling becomes confining at a larger mass
scale.  This is confirmed by our numerical estimates.  We infer a
confinement scale around $\mathcal{O}(\text{TeV})$, in order of magnitude.  We will refer
to $SO(5)$ as hypercolor, and the $SO(5)$ vector fermions as
hyperquarks.

$SO(5)$ supports a $Z_2$ conserved quantum number, which counts the
number of vector indices~\cite{Antipin:2015xia}.  It is analogous to
quark number (or baryon number) in QCD, but of course the distinction
between $Z_2$ and conventional, additive baryon number has major
physical consequences.  The lightest unconfined $Z_2$ odd ($SO(5)$
singlet) states are hyperbaryons.  In quark model language, they are
formed from 5 hyperquarks; in operator language, the lowest mass
dimension operator that creates them involves the product of 5
hyperquark fields.  Although they are highly stable individually,
hyperbaryons can annihilate into ordinary matter in pairs.
Conversely, they might be pair-produced in high energy collisions.

At high enough temperatures in the early universe, $T \gg 10$
  TeV, hyperbaryons would be in thermal equilibrium and their number
  density will be comparable to the photon number density.  As the
  temperature cools below their mass $M\sim 10$ TeV, their equilibrium abundance
  will diminish, until they become so rare that annihilation cannot
  keep up with the expansion of the universe, and a residual abundance
  freezes out. This scenario has a long history in cosmology.

  The ratio of the residual number density of hyperbaryons to photons
  is of order $\sim M/M_{\rm Planck}$, and the freezout temperature is
  parametrically less than $M$ by a logarithmic factor, roughly
  $\ln M/M_{\rm Planck}$.
  A more careful calculation, following~\cite{Griest:1989wd},
  gives
\begin{equation}
\Omega_\chi h^2 \approx 10^{-5}\Big( M/TeV\Big)^2
 \end{equation}

 Thus for $M \lesssim10$ TeV the relic hyperbaryons contribute only a
 small fraction of the mass density of the universe.  In consequence,
 though the current hyperbaryon relic abundance presents no obvious
 phenomenological catastrophe, the relic hyperbaryons might
 conceivably be detectable. One may also envisage
   that the lightest hyperbaryon might contribute to the dark matter
   density, as suggested in Ref. \cite{Mitridate:2017oky}.
   
 It is noteworthy that this cosmological mass bound ensures that if
 they exist at all, hyperbaryons are not far beyond the reach of
 high-energy accelerators currently under discussion.


\section{Summary and outlook} 

We have presented a model of comprehensive unification, bringing
together both gauge and family structure, with several attractive
features.  Within this approach, the existence of multiple fermion
families and the fact that they appear in spinor representations of
$SO(10)$ are intimately connected.  By combining orbifold projection,
Higgs symmetry breaking, and hypercolor confinement in a reasonably
simple way we can obtain just three chiral families, as is observed.
An interesting consequence is the emergence of highly stable
hyperbaryons, with mass $\sim$ 10 TeV, protected by a discrete $Z_2$
symmetry associated with the $SO(5)$ hypercolor group.
They provide an attractive target for accelerator and
astrophysical searches.
Finally, let us mention that one might attempt to pursue spinor
unification further, to bring in the space-time spinor structure, as
recently discussed in ~Ref.\cite{BenTov:2015gra}.

\section*{Acknowledgements}

Work supported by Spanish grants FPA2014-58183-P, Multidark
CSD2009-00064, SEV-2014-0398 (MINECO), PROMETEOII/2014/084
(Generalitat Valenciana). M.R. would like to thank Paula S\'aez for motivation during the initial stages of this work. C.A.V-A. acknowledges support form Mexican grant CONACYT No. 274397.  FW's work is supported by the U.S. Department of Energy under grant Contract  Number DE-SC0012567 and by the Swedish Research Council under Contract No. 335-2014-7424.

\bibliographystyle{apsrev} \providecommand{\url}[1]{\texttt{#1}}
\providecommand{\urlprefix}{URL }
\providecommand{\eprint}[2][]{\url{#2}}

\end{document}